\documentclass[amsmath,amssymb,showpacs,floatfix,twocolumn,pra,superscriptaddress]{revtex4}
\usepackage{graphicx}
\usepackage{dcolumn}
\usepackage{bm}

\def\beq{\begin{equation}}
\def\eeq{\end{equation}}

\newcommand{\bk}{\mathbf{k}}

\newcommand{\bq}{\mathbf{q}}
\newcommand{\br}{\mathbf{r}}

\newcommand{\la}{\langle}
\newcommand{\ra}{\rangle}

\newcommand{\eqname}[1]{\label{eq:#1}}
\newcommand{\ket}[1]{| #1 \rangle}

\newcommand\ek{\varepsilon_{\bf k}}
\newcommand\ekqp{\varepsilon_{\bf k}^{\rm{qp}}}
\newcommand\freq{\hbar\nu}
\newcommand\freqR{\hbar\omega-\varepsilon_3^o}
\newcommand{\gammak}{\Gamma_{\bf k}}
\newcommand{\Zk}{Z_{\bf k}}

\begin{document}

\title{Thermometry and signatures of strong correlations from Raman
spectroscopy of fermionic atoms in optical lattices}

\author{Jean-S\'ebastien Bernier}
\affiliation{Centre de Physique Th{\'e}orique,
\'Ecole Polytechnique, CNRS, 91128 Palaiseau Cedex, France}
\author{Tung-Lam Dao}
\affiliation{Laboratoire Charles Fabry de l'Institut d'Optique,
CNRS, Universit\'e Paris-Sud, Campus de l'{\'E}cole Polytechnique, 91127 Palaiseau Cedex, France}
\author{Corinna Kollath}
\affiliation{Centre de Physique Th{\'e}orique,
\'Ecole Polytechnique, CNRS, 91128 Palaiseau Cedex, France}
\author{Antoine Georges}
\affiliation{Centre de Physique Th{\'e}orique,
\'Ecole Polytechnique, CNRS, 91128 Palaiseau Cedex, France}
\affiliation{Coll\`ege de France, 11 place Marcelin Berthelot, 75005 Paris, France}
\author{Pablo S. Cornaglia}
\affiliation{Centro At\'omico Bariloche and Instituto Balseiro, CNEA, CONICET, 8400 Bariloche, Argentina}

\begin{abstract}
We propose a method to directly measure the temperature of a gas of weakly interacting
fermionic atoms loaded into an optical lattice. This technique relies on Raman spectroscopy and is applicable to
experimentally relevant temperature regimes. Additionally, we show that a similar spectroscopy scheme
can be used to obtain information on the quasiparticle properties and Hubbard bands of the metallic and Mott-insulating
states of interacting fermionic spin mixtures. These two methods provide experimentalists with novel probes to
accurately characterize fermionic quantum gases confined to optical lattices.
\end{abstract}

\date{\today}


\maketitle
\section{Introduction}

Fermionic ultracold atom physics has witnessed unprecedented experimental progress
since a quantum degenerate Fermi gas was first prepared in a three dimensional
optical lattice \cite{KoehlEsslinger2005}. The recent evidence for a fermionic 
Mott insulating state \cite{JoerdensEsslinger2008,SchneiderRosch2008}
serves as a clear example of these rapid advances. However, despite all
these breakthroughs, conducting experiments with fermionic atoms still remains a major
challenge as very few probes are available to accurately characterize these systems.
Among all difficulties encountered by experimentalists, the lack of reliable methods
to adequately measure the temperature of Fermi gases confined to optical lattices is often
cited as one of the main obstacles.

Although this problem is quite persistent when confronted to lattice systems,
the situation is much better in the continuum.
There, several techniques were successfully implemented to
estimate the temperature of both fermionic and bosonic quantum gases \cite{KetterleZwierlein2007,BlochZwerger2008, StewartJin2008}.
In the presence of an optical lattice, different schemes to determine the temperature have been devised and experimentally tested for bosonic atoms.
One approach relies on the direct comparison of experimental
and theoretical time-of-flight images obtained from computationally expensive simulations \cite{TrotzkyTroyer2009}.
In a second method, the temperature is estimated from a measurement of the width of the transition layer between
two spin domains created by the application of a magnetic field gradient \cite{WeldKetterle2009}. Relying on the good
local resolution attainable in two dimensional systems, a third approach
extracts temperature from the density and density fluctuations \cite{HungChin2009}. This
method, based on a generalized version of the fluctuation-dissipation theorem was analyzed
theoretically in \cite{ZhouHo2009}. Finally, detecting the temperature using  bosonic impurity atoms which are insensitive
to the optical lattice potential was put forward in \cite{DeMarco2009}.

The situation is much more difficult for fermionic atoms loaded into an optical lattice.
In current experiments, temperature measurements are usually performed before switching on the lattice
potential and after switching it off \cite{JoerdensEsslinger2008,SchneiderRosch2008}. However,
since the temperature changes during the loading process, detecting the system temperature
with the optical lattice on is of the utmost importance. Experimental attempts
\cite{StoeferleEsslinger2006} 
at evaluating the temperature of fermions inside a lattice were based on
the determination of the number of doubly occupied sites.
In this scheme, an accurate evaluation of the temperature requires a full theoretical
understanding of the strong dependence of the number of doubly occupied sites with temperature \cite{Koehl2006,KatzgraberTroyer2006,DeLeoParcollet2008}.
In addition, a precise experimental knowledge of the interaction strength, hopping amplitude, trapping configuration
and particle number is needed. The combination of all these requirements renders this approach difficult to
use. On the theoretical side, other methods were proposed. 
For example, one could envisage, as proposed in \cite{ZhouHo2009}, to extract
temperature from a generalized version of the fluctutation-dissipation theorem used in conjunction
with the knowledge of both spatially resolved system density and density fluctuations. However,
this approach needs sufficiently strong density fluctuations as well as very good local resolution,
an experimental requirement that is far from being met for fermionic systems. Finally, in one and two
dimensions, measuring the intensity of the light scattered off the atomic lattice array was
proposed to detect the system temperature \cite{RuostekoskiDeb2009}.

In this work, we propose a novel method to measure the temperature of fermions
loaded into two or three dimensional optical lattices. The approach we put forward relies on
transferring a portion of the atoms stored into the optical lattice potential to a
third hyperfine state using a stimulated Raman process.
Hence, the temperature measurement can either be done locally or globally, and can be used in
parallel with other probes. Depending on the experimental resolution, this thermometer works for
both free and weakly interacting fermions, and only requires the knowledge of the hopping amplitude
of the system under study. In addition, our approach can be implemented using present fermionic
ultracold atom technology.

Measuring the temperature of fermionic gases loaded into optical lattices is not the
only difficulty faced by experimentalists working with cold atoms. Indeed, identifying the different
strongly correlated phases that can be realized in these systems is also a demanding task. In
comparison to condensed matter systems, few probing techniques are available to study cold atomic
systems. Spectroscopic methods \cite{BlochZwerger2008, KetterleZwierlein2007}, including momentum-resolved 
radio-frequency spectroscopy \cite{StewartJin2008}, have been shown previously to be very efficient in probing characteristics of
quantum gases not subjected to optical lattice potentials.
Recently, it has been suggested that Raman spectroscopy can also be used to probe the excitation spectrum
of strongly correlated phases of Bose gases confined to optical
lattices \cite{KonabeNakamura2006, Blakie2006} and
to investigate single-particle excitations in normal and superfluid phases of
fermionic gases \cite{DaoCarusotto2006, DaoGeorges2009, Dao2008}.
In this article, we show that spectroscopy can be used to identify various signatures
of strongly correlated fermionic phases in optical lattices. Experimentally detectable features
include the presence of quasiparticle peaks in weakly and strongly correlated liquids
as well as Hubbard bands in strongly correlated liquids and Mott insulators.

The rest of the article is organized as follows: in Sec. \ref{sec:setup} we define the general setup
for Raman spectroscopy. In Sec. \ref{sec:temp} we present the temperature detection scheme for non-interacting
(Sec. \ref{sec:noninteracting}) and weakly interacting (Sec. \ref{sec:weakly}) Fermi gases confined to two- and
three-dimensional optical lattice. Finally, in Sec. \ref{sec:Mott} we demonstrate that spectroscopy
can also be used to identify various signatures of strongly correlated systems.

\section{Setup and theoretical description}
\label{sec:setup}

\begin{figure}[!ht]
  \includegraphics[width=0.7\linewidth,clip=true]{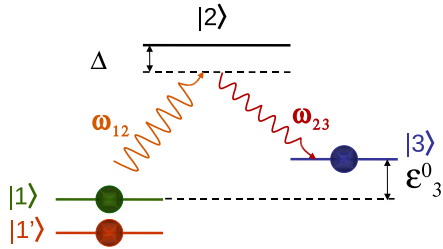}
  \caption{Atomic levels involved in the Raman process. Atoms in hyperfine
          state $\ket{1}$ are transferred to state $\ket{3}$ using two Raman laser beams
          with frequencies $\omega_{12}$ and $\omega_{23}$, respectively. These frequencies are both
          detuned by $\Delta$ from their corresponding resonances to state $\ket{2}$. States
          $\ket{1}$ and $\ket{3}$ are separated in energy by $\varepsilon^{o}_3$. \label{fig:setup_raman}}
\end{figure}

\begin{figure}[!ht]
 \includegraphics[width=0.7\linewidth,clip=true]{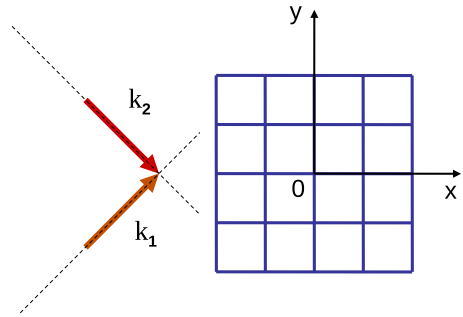}
  \caption{Geometrical configuration of the two Raman laser beams carrying
           momentum ${\bf k}_1$ and ${\bf k}_2$ respectively.\label{fig:setup_angle}}
\end{figure}

The proposed detection schemes rely on Raman spectroscopy \cite{DaoCarusotto2006, DaoGeorges2009, FootNote1}. 
This probing technique consists in exciting with a given energy and momentum
a many-body state formed of a mixture of two hyperfine states by transferring atoms to a
third state. To set the ideas straight, we sketch the Raman
process in Fig.~\ref{fig:setup_raman}. There we see that atoms from hyperfine state $\ket{1}$
are transferred to a different hyperfine state $\ket{3}$ using two Raman laser beams with
frequencies $\omega_{12}$ and $\omega_{23}$ and Rabi frequencies $\Omega_{12}$ and $\Omega_{23}$, respectively.
The frequencies $\omega_{12}$ and $\omega_{23}$ are both detuned from their corresponding resonances to
state $\ket{2}$ to keep this state unoccupied. During the transition,
momentum ${\bf q}={\bf k}_1-{\bf k}_2$ is transferred to the atoms. This value can
be chosen within certain bounds by adjusting  appropriately the angles between the two Raman beams and
the lattice axis (Fig.~\ref{fig:setup_angle}). For example, transferred momentum $\bq=0$ could 
be realized using copropagating Raman laser beams. 
While, for many current lattice setups, 
$\bq \approx (\frac{\pi}{a},\frac{\pi}{a})$ or $(\frac{\pi}{a},\frac{\pi}{a},\frac{\pi}{a})$ 
could  be reached by aligning two counterpropagating Raman lasers along the diagonal of the optical lattice axes.

Experimentally, the Raman signal is measured
by counting the number of atoms transferred to state $\ket{3}$. This signal can in principle
be resolved both in frequency and  momentum. For many applications, such as thermometry, momentum 
resolved measurements are not needed. Nevertheless, as we will show in Sec. \ref{sec:Mott}, momentum resolution 
can also provide valuable additional information, but achieving good momentum resolution in optical 
lattice setups is experimentally demanding. 

When only a small fraction of the atoms in hyperfine state $\ket{1}$ are
transferred into state $\ket{3}$, the Raman signal can usually be approximated
using a linear response expression \cite{DaoCarusotto2006, DaoGeorges2009}.
Within local density approximation, the Raman transition rate is given by
\begin{eqnarray}
R_\bq(\omega) &=& \frac{2\pi}{\hbar} \sum_\br \int d{\bf k}~W_{\bf k}^{\bf q}~
|\Omega_e(\br)|^2~n_F(\varepsilon^{\bf r}_{3,{\bf k}}-\hbar\omega-\mu_o) \nonumber \\
&& ~~~~~~ \times A({\bf k}-{\bf q},\varepsilon^{\bf r}_{3,{\bf k}} - \mu_o - \hbar\omega; \mu_{\bf r}).
\label{eq:rate}
\end{eqnarray}
In this expression, $\varepsilon^{\bf r}_{3,{\bf k}} = \varepsilon_{3,{\bf k}} + V_{3}({\bf r})$ and
$\mu_{\bf r} = \mu_o - V_{1}({\bf r})$ where $\varepsilon_{3,{\bf k}}$ is the dispersion relation for
the $\ket{3}$ state, $V_{1,3}({\bf r})$ are the trapping potentials felt by states $\ket{1}$ and
$\ket{3}$, respectively, and $\mu_o$ is the chemical potential in the center of the trap. The momentum dependent coefficient
$W_{\bf k}^{\bf q}$ is due to the Wannier envelope and is given by
\begin{eqnarray}
W_{\bf k}^{\bf q} &=&~|~\int d{\bf r}~w^*_{1}({\bf r})~\psi_{3,{\bf k}}({\bf r})~
e^{-i{\bf q}\cdot{\bf r}}~|^2\\ \nonumber
&=&~|~\int d{\bf r}\, u^*_{1,\bf k}({\bf r})\,u_{3,\bf{k-q}}({\bf r})~|^2
\end{eqnarray}
where $w_{1}({\bf r})$ is the Wannier function for the atoms in state $\ket{1}$ while $\psi_{3,{\bf k}}({\bf r})$
is the Bloch function for the atoms in state $\ket{3}$
($u_{1,\bf k},u_{3,\bf k}$ are the corresponding periodic parts of the Bloch function).
$n_F(x) = 1/(1+\exp(x/(k_BT)))$ is the
Fermi function. $A({\bf k},\hbar\nu;\mu_r)$ is the one-particle spectral function for the
$(\ket{1},\ket{1'})$ mixture in a confining potential. The local density approximation has been used, so that
$\mu_\br$ is the local chemical potential at point ${\bf r}$. 
In an homogeneous system, the spectral function is defined as
\begin{multline}
A(\bk,\hbar\nu)=\sum_{i,f} \frac{e^{-\bar{E}_i/k_B T}+e^{-\bar{E}_f/k_B
T}}{\mathcal{Z}}\\
\times
\left|\la\phi_{f}|c_{1\bk}|\phi_{i}\ra\right|^2\,\delta(\hbar\nu+\bar{E}_f-\bar{E}_i)
\eqname{spectral_homo}
\end{multline}
where $c_{1\bk}$ detroys an atom in state $\ket{1}$ with momentum ${\bf k}$,
${\mathcal{Z}} = \sum_i\exp(-\bar{E}_i/k_BT)$ is the Grand-Canonical partition function, the
sums over $i$ and $f$ refer to all the many-body states of the system and the energy
$\bar{E}_i = E_i - \mu N_i$ is rescaled by the number of particles. Finally,
$\hbar\omega = \hbar(\omega_{12} - \omega_{23})$ is the transferred energy, and the Rabi frequency
is $ \Omega_e(\br)= \Omega_{12}(\br) \Omega_{23}^{*}(\br)/\Delta$, where $\Delta$ is the detuning.
Local resolution
of the Raman transfer could be obtained by using special configuration of laser
beams \cite{DaoGeorges2009}.

From Eq.~\ref{eq:rate}, we see that for a given position and momentum the Raman spectrum
is obtained from the multiplication of two functions. The first function is the Fermi factor $n_F$
which depends strongly on temperature, but is independent of other parameters apart
from $\mu_o$ and $T$ itself.
As we will show in Sec. \ref{sec:temp}, our temperature detection scheme relies primarily on this
observation. The second function entering Eq.~\ref{eq:rate} is the spectral function which
depends sensitively on the state of the system. We will see that this
limits somewhat the possibility of a universal temperature determination,
but as we explain in Sec. \ref{sec:Mott} this function provides valuable
information on the phase of the system.

Through out the rest of this article, we describe quantum gases confined to optical lattice
potentials using the fermionic Hubbard model \cite{JakschZoller1998,HofstetterLukin2002}:
\begin{eqnarray}
   \label{eq:h}\nonumber
   H =
   -J \sum_{\langle r,r'\rangle\sigma} \left(c_{r\sigma}^\dagger
   c^{\phantom{\dagger}}_{r'\sigma}+h.c.\right)
   &+& U \sum _{r} \hat{n}_{r\uparrow} \hat{n}_{r\downarrow}
   \nonumber \\ 
   &-& \sum_{r\sigma}\,\mu_r~\hat{n}_{r\sigma},
\end{eqnarray}
where $c^\dagger_{r\sigma}$ and $c_{r\sigma}$ are the creation and annihilation operators of
the fermions with $\sigma = \{\ket{1},\ket{1'}\}$,
$J$ is the hopping matrix element, $U$ is the on-site repulsion, $\mu_r$ is the local chemical potential
and $ \hat{n}_{r\sigma}= c^\dagger_{r\sigma} c^{\phantom{\dagger}}_{r\sigma}$ is the number operator on site $r$. $\langle r,r'\rangle$ denotes neighboring lattice sites. 


\section{Temperature determination}
\label{sec:temp}

In this section, we present a novel method to evaluate the temperature of weakly interacting fermions
confined to optical lattices by measuring their Raman spectrum. We first present this method
considering non-interacting fermions in two- and three-dimensional optical lattices. Afterwards, we
demonstrate how this procedure is also applicable to weakly interacting fermionic gases
loaded into optical lattices.

\subsection{Temperature extraction for non-interacting fermions}
\label{sec:noninteracting}
For the case of non-interacting fermions, i.e. $U = 0$, the spectral function
for trapped $\ket{1}$ atoms is $A(\bk,\hbar\nu;\mu_r)=\delta(\hbar\nu + \mu_{r}- \varepsilon_{1,{\bf k}})$.
In our study, we will assume that both $\ket{1}$ and $\ket{3}$ atoms are trapped by the same harmonic
potential, i.e.~$V_1 = V_3 = V_T$, and are loaded into an optical lattice which is felt equally by both hyperfine states, 
i.e.~induces the same hopping coefficients, 
such that $\varepsilon_{1,{\bf k}} = \varepsilon_{3,{\bf k}} - \varepsilon_3^o \equiv \varepsilon_{{\bf k}}$.
Here $\varepsilon_3^o$ is the energy offset of state $\ket{3}$ with respect to state $\ket{1}$ (cf.~Fig.~\ref{fig:setup_raman}).
These assumptions are valid as long as we use adequate hyperfine states and confine the
atoms into far-detuned optical lattices. Under these conditions, the Raman transition rate is given by
\begin{eqnarray}
R_\bq(\omega) &=&~\frac{2\pi}{\hbar}~\sum_r~\int~d{\bf k}~W^{\bf q}_{\bf k}~|\Omega_e(\br)|^2 \nonumber \\
&& ~~\times n_F(\varepsilon_3^o -\hbar\omega + \varepsilon_{{\bf k}} + V_T({\bf r}) -\mu_o) \nonumber \\
&& ~~\times \delta(\varepsilon_3^o - \hbar\omega + \varepsilon_{{\bf k}} - \varepsilon_{{\bf k-q}}).
\label{eq:lattice}
\end{eqnarray}

\begin{figure}[!ht]
\includegraphics[width=0.45\linewidth,clip=true]{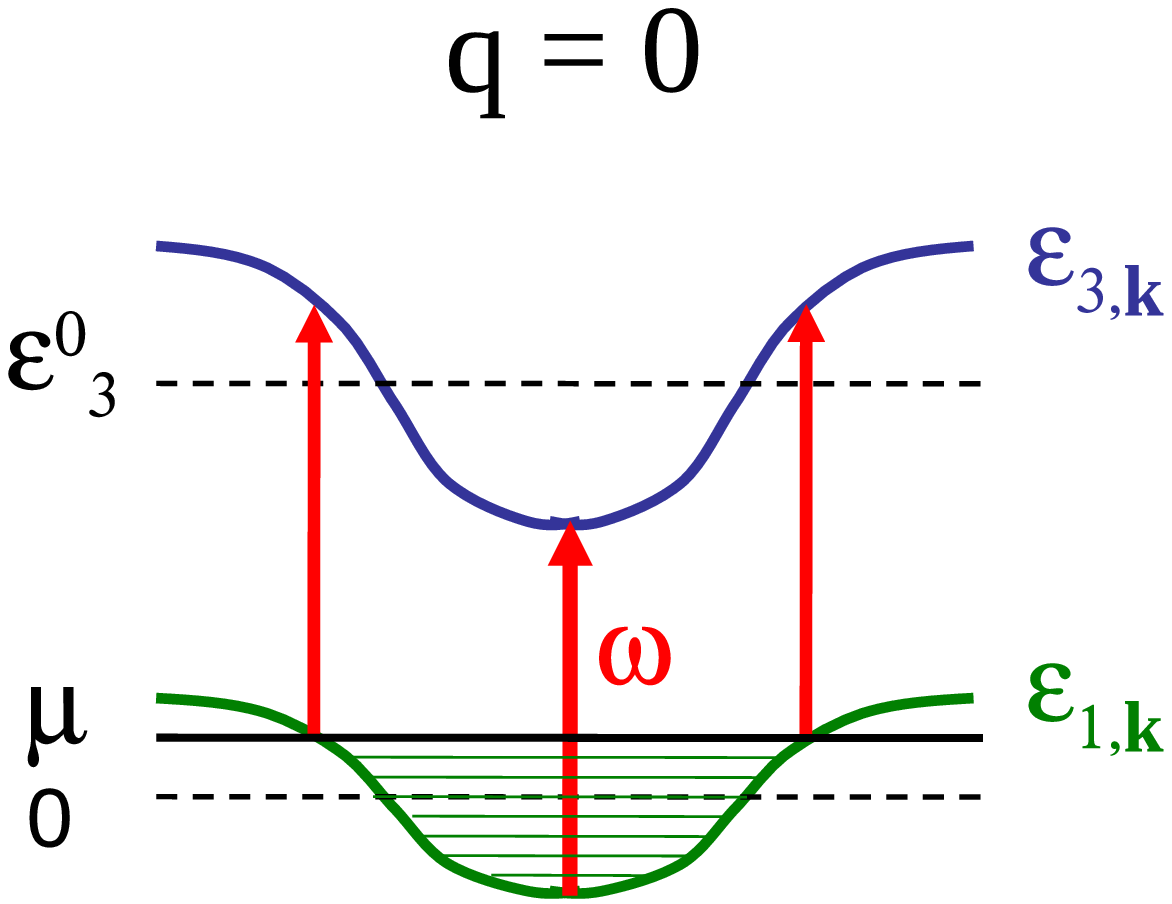}
\includegraphics[width=0.45\linewidth,clip=true]{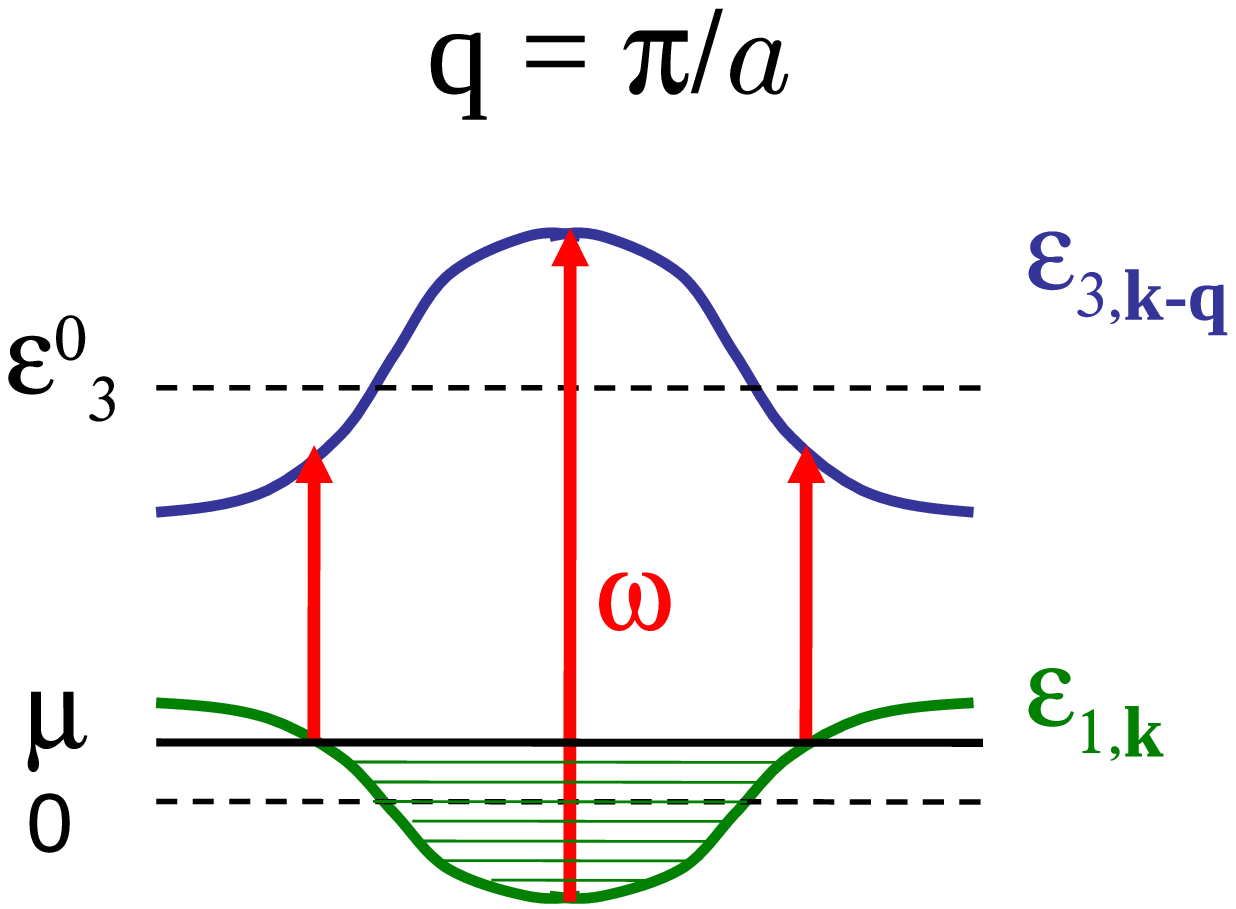}
  \caption{Left: Raman transfer with ${\bf q} = 0$, atoms can only be transferred from states $\ket{1}$ to
           $\ket{3}$ with frequency $\omega = \varepsilon_3^o/\hbar$. Right: Raman transfer with
           ${\bf q} = \pi/a$, atoms can in principle be transferred from states $\ket{1}$ to $\ket{3}$
           with frequencies ranging from $(\varepsilon_{3}^{o}-2D)/\hbar$ to $(\varepsilon_{3}^{o}+2D)/\hbar$ where $D$ is
           the half-bandwidth. $\varepsilon_{1,{\bf k}}$ and $\varepsilon_{3,{\bf k}}$
           are the dispersions of states $\ket{1}$ and $\ket{3}$, respectively, $\mu$ is the chemical potential
           for the mixture of states $\ket{1}$ and $\ket{1'}$, and $\varepsilon_{3}^{o}$ is the energy shift of
           $\ket{3}$ compared to $\ket{1}$.\label{fig:hom_bands}}
\end{figure}

This expression depends on temperature only through the Fermi function $n_F$. 
Therefore, detecting the system temperature can be done
reliably by fitting $R_{\bf q}(\omega)$ with a minimum of parameters.
From Eq.~\ref{eq:lattice}, one also sees that the frequency spread of the Raman signal is
strongly dependent on the chosen transferred momentum, ${\bf q}$. As shown on
Fig.~\ref{fig:hom_bands}, for a homogeneous system, atoms can only be
transferred from states $\ket{1}$ to $\ket{3}$ with energy $\hbar\omega = \varepsilon_3^o$
if ${\bf q} = 0$.
Hence, $R_{{\bf q}=0}(\omega)$ is strongly peaked at $\varepsilon_3^o/\hbar$ and zero everywhere
else
\cite{FootNote2}. 
This feature makes it impossible to detect the system temperature using ${\bf q} = 0$
as the Raman signal is too narrow in $\omega$. What we need is a Raman signal that is non-zero for a wide
range of frequencies. This is achieved if ${\bf q} = (\frac{\pi}{a},\frac{\pi}{a})$ for square lattices and
${\bf q} = (\frac{\pi}{a},\frac{\pi}{a},\frac{\pi}{a})$ for cubic lattices.
As we can see on Fig.~\ref{fig:hom_bands}, for these two configurations,
since $\varepsilon_{\bf k-q} = - \varepsilon_{\bf k}$, atoms can in principle be transferred
with energies ranging from $\varepsilon_{3}^{o}-2D$ to $\varepsilon_{3}^{o}+2D$ where $D$ is the half-bandwidth.
Therefore, temperature measurements can be done with this choice of transferred momentum.
We use the notation ${\bf q} = \pi/a$ to specify that we use these configurations. We also note that since we
are interested in the momentum integrated Raman rate, $W^{{\bf q}=\pi/a}_{\bf k}$ can be neglected without loss of
generality for sufficiently deep lattices as this factor only affects the signal overall amplitude \cite{FootNote3}.


In the remaining of this section, we demonstrate that Raman spectroscopy can adequately be used
as a thermometer. We conduct this demonstration in two steps. First, using Eq.~\ref{eq:lattice}, we
numerically simulate a Raman experiment in which the probing lasers are shone on the whole system of
atoms confined to a lattice and trapped in an harmonic
potential. These simulations show that the resulting spectra vary significantly with temperature.
Then, as a second step, we fit the obtained signal using two different fitting functions.
The first is the full continuum-space expression for the Raman rate
(valid for large systems):
\begin{eqnarray}
\frac{R_{\pi/a}(\tilde{\omega})}{N} \propto \frac{C}{\rho}~g_v\left(\frac{\hbar\tilde{\omega}}{2}\right)
\int_{-\infty}^{\mu_o} \! d \mu \,\frac{(\mu_o-\mu)^{(d-2)/2}}{1+e^{-(\hbar\tilde{\omega}/2+\mu)/(k_BT)}}.
\label{eq:rate_free}
\end{eqnarray}
In this expression, $\hbar \tilde{\omega}= \hbar \omega-\varepsilon^o_{3}$,
$g_v(\varepsilon)\equiv \frac{1}{V} \sum_\bk\delta(\varepsilon-\varepsilon_\bk)$ is the density
of states of the band, $d$ is the dimension of the system, $V$ its volume and
$C = 2\pi|\Omega_e|^2/\hbar$. 
As we use the local 
density approximation, the Raman rate per particle only depends on the particle number and the trapping potential
through the characteristic particle number $\rho=N (V_T/D)^{d/2}$ where $N$ is the total
number of atoms in the hyperfine mixture $(\ket{1}, \ket{1'})$. We are left with only three fitting
parameters: the temperature, $T$, the chemical potential at the center of the trap, $\mu_o$, and
an overall multiplicative factor. The second fitting function is the simplified expression:
\begin{eqnarray}
R_{\pi/a}(\tilde{\omega})\propto g_v(\hbar \tilde{\omega}/2) e^{(\hbar\tilde{\omega}/2k_BT)}.
\end{eqnarray}
As we will show this approximate expression which has only temperature and a prefactor as fitting parameters is
only valid in certain parameter regimes. However, it has the advantage of being very simple.

In the following, we show that the values of temperature and
central chemical potential obtained through this fitting procedure agree very well with the
initial system parameters. This means that each Raman spectrum is to a good extend uniquely defined by its temperature
and chemical potential both in two and three dimensions.

\subsubsection{Non-interacting fermions in two dimensions}

\begin{figure}[!ht]
  \includegraphics[width=0.85\linewidth,clip=true]{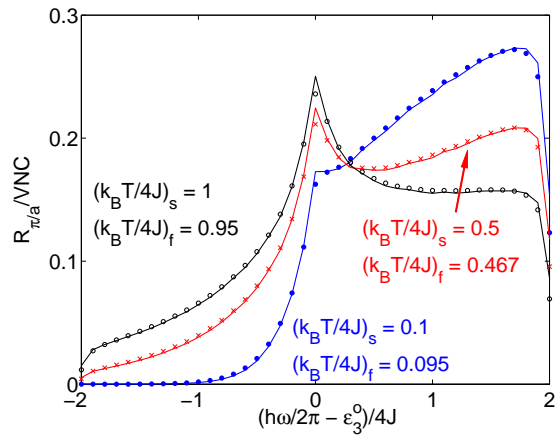}
  \caption{Raman spectra for three different temperatures $k_B T/4J = 0.1, 0.5,1$ at the
  characteristic particle number $\rho = 2$ in a two-dimensional system. $(k_B T/4J)_{s}$ is the exact temperature
  while $(k_B T/4J)_{f}$ is obtained by fitting the spectra to Eq.~\ref{eq:rate_free_2D}.\label{fig:2D_diffT_samerho}}
\end{figure}

Let us first look at Raman spectra for a two-dimensional system with fixed characteristic particle number $\rho = 2$ and varying
temperatures. These spectra are shown in Fig.~\ref{fig:2D_diffT_samerho}. From this figure, one
can see that all spectra can be broken down into two parts. For large $\tilde{\omega}$,
each spectrum is characterized by a signal of large amplitude whose shape depends on the system
parameters while, for small $\tilde{\omega}$, each spectrum presents a tail whose shape
is mainly set by temperature. The peak or discontinuity at $\tilde{\omega} = 0$ is due to the 
Van Hove singularity of the square lattice density of states. The presence of sharp
edges at $\hbar \tilde{\omega}/4J= \{-2, 2\} $ is due to the abrupt ends of the square lattice density of
states. At low temperatures, most of the spectral weight is
located at $\tilde{\omega} > 0$ with a sharp step around  $\tilde{\omega}\approx 0$ (cf. $k_BT/4J=0.1$).
As the system temperature is increased, some weight is transferred into the tail and the step broadens
considerably. This broadening stems from the smoothening of the Fermi function with
increasing temperature. For high temperatures, the left end of the spectrum becomes sharp as it is cut 
by the edge of the density of states.

\begin{figure}[!ht]
 \includegraphics[width=0.85\linewidth,clip=true]{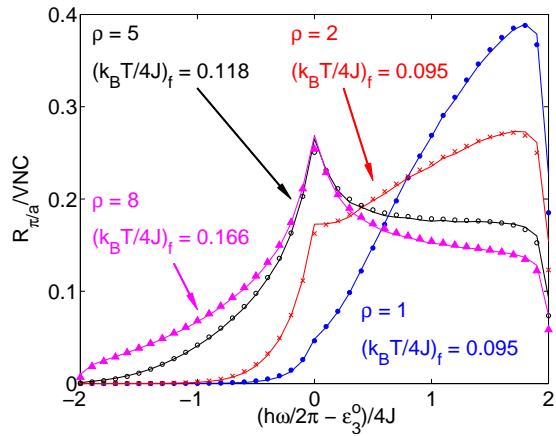}
  \caption{Raman spectra for four different characteristic particle numbers $\rho = 1, 2, 5, 8$ at fixed
           temperature $k_B T/4J = 0.1$ in a two-dimensional system.
           $(k_B T/4J)_{f}$ is obtained by fitting the spectra to
           Eq.~\ref{eq:rate_free_2D}. Note that the case where the fit does not provide an accurate
           determination ($\rho=8$) corresponds to an almost filled band at the trap center, with few available thermally
           excited states.\label{fig:2D_sameT_diffrho}}
\end{figure}

In Fig.~\ref{fig:2D_sameT_diffrho}, we show the dependence of Raman spectra with varying
characteristic particle numbers at a fixed temperature. For small $\rho$, most of the weight is
located in the bulk of the spectrum ($\tilde{\omega} > 0$), whereas, at higher $\rho$, more and more
weight shifts into the tail. For very large characteristic particle number,
the left edge of the spectrum is located at the end of the $\tilde{\omega}$ window allowed
by the support of the density of states and not where the Fermi function goes to zero.

\begin{figure}[ht]
\includegraphics[width=0.85\linewidth,clip=true]{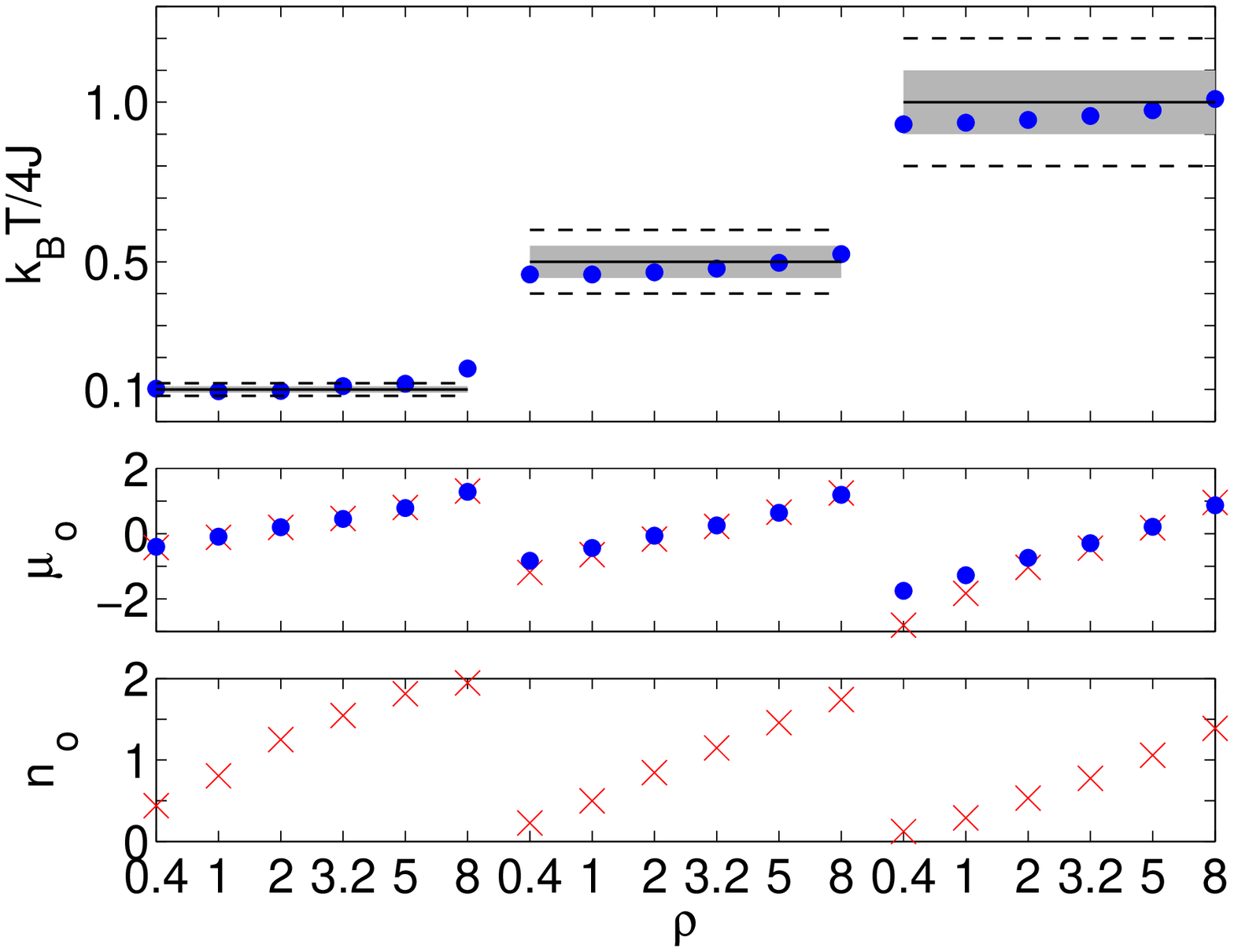}
\caption{Accuracy of detected temperatures and central chemical
         potentials. These detected values were obtained by fitting each full spectrum with Eq.~\ref{eq:rate_free_2D}.
         Upper panel: detected temperatures are denoted by blue dots while
         exact temperatures by solid black lines. Each shaded region corresponds to a $10\%$ range centered on the
         exact temperature and contains most fitted points. The regions delimited by dashed lines corresponds
         to a $20\%$ range. Central panel: central chemical potentials are detoned by blue dots, exact values by
         red ``$X$''. Lower panel: density at the center of the trap $n_0$.
\label{fig:quality_anal_2D}}
\end{figure}

From the above descriptions, it is clear that the Raman spectra strongly depend
on temperature and particle density. We exploit this strong dependence by fitting
these Raman spectra to an integrated version of Eq.~\ref{eq:rate_free}:
\begin{eqnarray}
\frac{R^{2D}_{\pi/a}(\tilde{\omega})}{N} \propto \frac{T}{\rho}~g_v(\hbar \tilde{\omega}/2)~\ln(1+e^{(\hbar\tilde{\omega}/2+\mu_o)/k_BT}).
\label{eq:rate_free_2D}
\end{eqnarray}
From this fit, we extract the temperature and central chemical potential of a two-dimensional
system. The values obtained agree very well with the initial system parameters.
We summarize the accuracy of the fitted parameters in Fig.~\ref{fig:quality_anal_2D}. In
the upper panel, we see that, asides from small deviations, the temperature can be determined very accurately by this
procedure and is in most cases well within $10\%$ of its true value. In the second panel, we show that the
chemical potential in the center of the trap can also be evaluated. Even though its accuracy is not as good as for
temperature, it still agrees within $20\%$ for most of the simulated systems. Therefore, from the knowledge
of experimentally measurable Raman spectra, we can accurately determine the temperature and obtain a good
estimate of the chemical potential at the
center of the trap for a free fermionic gas confined to a two dimensional optical lattice. Let us point out
that here we assumed that the hopping amplitude in the optical lattice, $J$, is known. However, as explained
earlier, $J$ sets the support of the spectra at high fillings and can therefore be experimentally detected as a bonus
by Raman spectroscopy measurements.

\begin{figure}[!ht]
 \includegraphics[width=0.85\linewidth,clip=true]{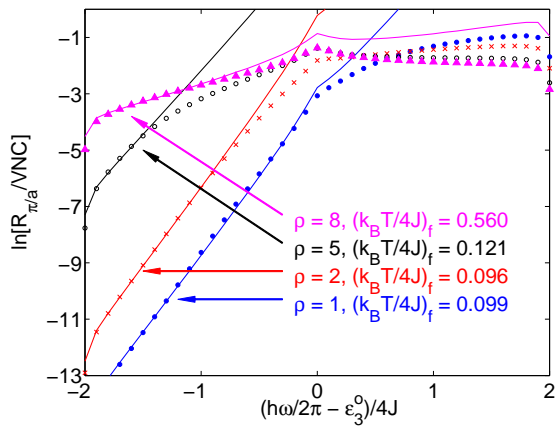}
  \caption{Raman spectra in logarithmic scale for four different characteristic particle numbers $\rho = 1, 2, 5, 8$ 
 at fixed
   temperature $k_B T/4J = 0.1$ in a two-dimensional system. $(k_B T/4J)_{f}$ is obtained by fitting the tails of the spectra to
   Eq.~\ref{eq:2D_slope}.\label{fig:2D_sameT_diffrho_log}}
\end{figure}

\begin{figure}[!ht]
 \includegraphics[width=0.85\linewidth,clip=true]{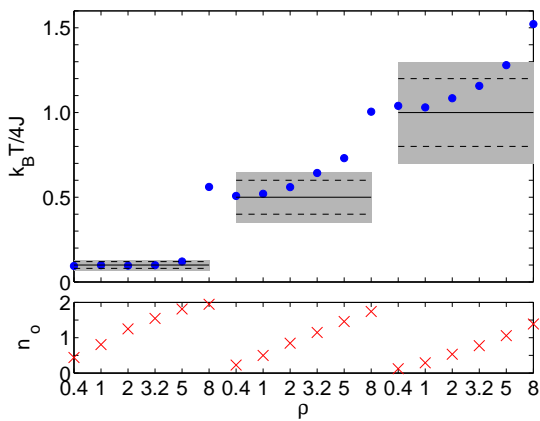}
  \caption{Accuracy of detected temperatures obtained by fitting the tail of each spectrum with Eq.~\ref{eq:2D_slope}.
           Upper panel: detected temperatures are denoted by blue dots while exact temperatures by solid black lines.
           Each shaded region corresponds to a $30\%$ range centered on the exact temperature and contains
           most fitted points. The regions delimited by dashed lines corresponds to a $20\%$ range.
           Lower panel: density at the center of the trap.\label{fig:quality_anal_2D_log}}
\end{figure}

In many cases a much simpler fitting procedure can already give very good results for the temperature.
This simplified method focuses on the behavior of the low frequency part of the Raman spectrum tail.
In Fig.~\ref{fig:2D_sameT_diffrho_log}, the Raman spectrum is shown on a logarithmic scale
in order to emphasize its tails. Looking back at the analytical expression given by Eq.~\ref{eq:rate_free_2D},
we see that if $e^{(\hbar\tilde{\omega}/2+\mu_o)/k_BT}$ is small the Raman signal can be approximated by
\begin{eqnarray}
R^{2D}_{\pi/a}(\tilde{\omega})\propto g_v(\hbar \tilde{\omega}/2) e^{(\hbar\tilde{\omega}/2k_BT)}.
\label{eq:2D_slope}
\end{eqnarray}
This simplified expression can be used as long as its validity extends over a sufficiently large region
of measurable signal. In other words, if $e^{(\hbar\tilde{\omega}/2+\mu_o)/k_BT} \ll 1$ for a wide range
of $\hbar\tilde{\omega}/4J \in [-2;2]$. This condition is most easily fulfilled for small or even
negative $\mu_o$. Hence, this simplified fitting procedure holds best for small and intermediate characteristic
densities. These limitations are quite apparent in Fig.~\ref{fig:2D_sameT_diffrho_log}. For example, increasing the
characteristic density increases the signal in the tail but drastically reduces the region over which the fitting
procedure works. For $\rho=5$, even at $k_BT/4J=0.1$, only a very small region is left. The same behavior is
also observed at larger temperatures. In fact, the good fitting region completely drops off the spectrum at
large temperatures and large characteristic densities as this region would appear 
outside $\hbar\tilde{\omega}/4J \in [-2;2]$, the range permitted by the
density of states.
The resulting accuracy of the simplified fitting procedure is summarized
for different system parameters in Fig.~\ref{fig:quality_anal_2D_log}. Here the simplified fit works very well for
the lowest values of $\rho$ whereas for larger characteristic densities the temperature is overestimated. 

\subsubsection{Non-interacting fermions in three dimensions}

\begin{figure}[!ht]
  \includegraphics[width=0.85\linewidth,clip=true]{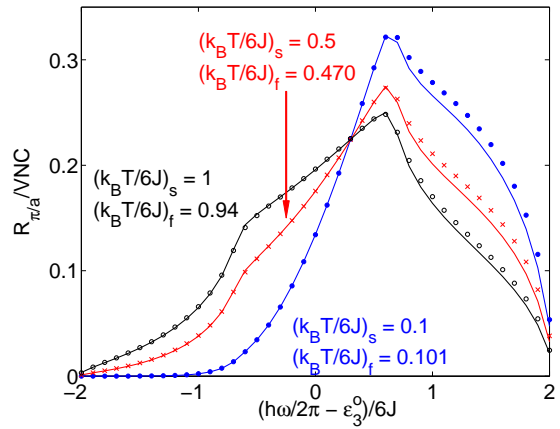}
  \caption{Raman spectra for three different temperatures $k_B T/6J = 0.1, 0.5,1$ at a fixed characteristic particle
           number $\rho=2.5$ in three dimensions. $(k_B T/6J)_{s}$ is the exact temperature
           while $(k_B T/6J)_{f}$ is obtained by fitting each spectrum up to its peak with Eq.~\ref{eq:rate_free3D}.\label{fig:3D_diffT_samerho}}
\end{figure}

\begin{figure}[!ht]
 \includegraphics[width=0.85\linewidth,clip=true]{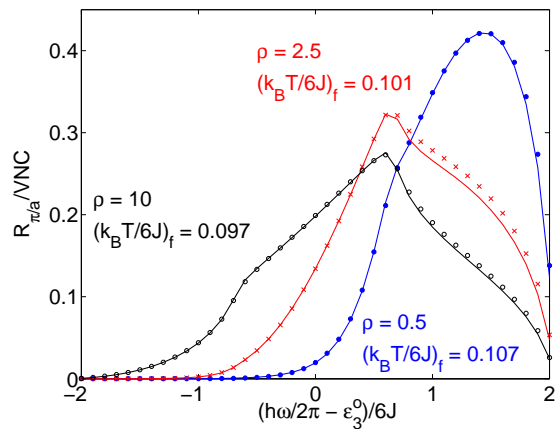}
  \caption{Raman spectra for three different characteristic particle numbers $\rho= 0.5, 2.5, 10$ at fixed temperature
           $k_B T/6J=0.1$ in three dimensions. $(k_B T/6J)_{f}$ is obtained by fitting each spectrum up to its peak with 
           Eq.~\ref{eq:rate_free3D}.\label{fig:3D_sameT_diffrho}}
\end{figure}

Let us now look at Raman spectra for atoms confined to a cubic lattice.
For fixed characteristic particle number and various temperatures, typical signals are shown in Fig.~\ref{fig:3D_diffT_samerho}.
At low temperatures, these spectra show a main peak and a tail that quickly goes to zero.
For increasing temperatures, the structure of the cubic density of states becomes more apparent as
weight shifts towards the tail. This change in the shape of the Raman spectrum is due to the
Fermi function whose spread increases with temperature. In Fig.~\ref{fig:3D_sameT_diffrho}
the evolution of the spectra with increasing characteristic particle number, $\rho$,
is shown at low temperature. For intermediate filling, the Raman spectrum presents
its characteristic tail whose size depends strongly on temperature. At very large
filling, the tail cannot be followed until its end as the Raman spectrum
is limited by the frequency window allowed by the cubic density of states.


\begin{figure}[ht]
\includegraphics[width=0.85\linewidth,clip=true]{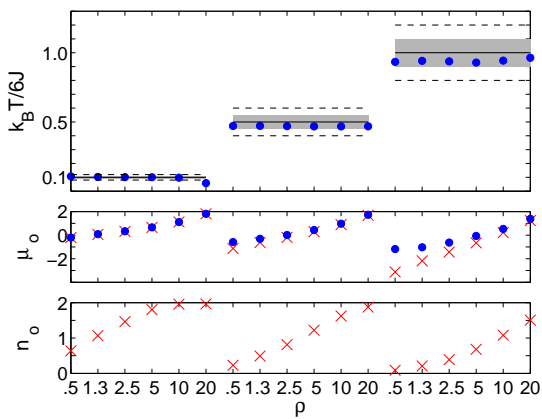}
  \caption{Accuracy of detected temperatures and central chemical
            potentials. These detected values were obtained by fitting each spectrum up to its peak with Eq.~\ref{eq:rate_free3D}.
            Upper panel: detected temperatures are denoted by blue dots while
            exact temperatures by solid black lines. Each shaded region corresponds to a $10\%$ range centered on the
            exact temperature and contains most fitted points. The regions delimited by dashed lines corresponds
            to a $20\%$ range. Central panel: detected central chemical potentials are detoned by blue dots, exact values by
            red ``$X$''. Lower panel: density at the center of the trap.
            \label{fig:quality3D}}
\end{figure}

In three dimensions, we can also extract the temperature from the measured signal.
In this case, the Raman rate in the continuum approximation is given by
\begin{eqnarray}
\frac{R^\text{3D}_{\pi/a}(\tilde{\omega})}{N}\propto \frac{g_v(\hbar \tilde{\omega}/2)}{\rho}
\int_{-\infty}^{\mu_o} \! d \mu \,\frac{(\mu_o-\mu)^{1/2}}{1+e^{-(\hbar\tilde{\omega}/2+\mu)/(k_BT)}}.
\label{eq:rate_free3D}
\end{eqnarray}
By fitting three dimensional spectra with Eq.~\ref{eq:rate_free3D}, we checked that the temperature and
central chemical potential are, to a good degree, uniquely defined for a given spectrum.
The quality of the extracted temperature and central chemical potential values are
summarized in Fig.~\ref{fig:quality3D}. We find very good agreement between the input and
extracted temperatures if the fit is done from $\tilde{\omega}_\text{min}$ to the $\tilde{\omega}$
value corresponding to the peak of the spectrum. However, fitting over the whole spectrum is
not as successful as more importance is given to the rightmost portion of the signal which is not as
sensitive to temperature as the tail is. Using the reduced fitting range, the
temperature can be determined within $10\%$ uncertainty. The chemical potential
at the center of the trap can also be determined. However, the agreement between the input
and extracted values decreases with increasing temperature.

\begin{figure}[!ht]
 \includegraphics[width=0.85\linewidth,clip=true]{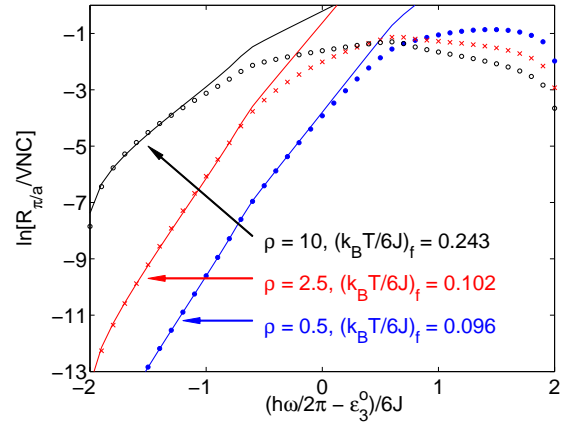}
  \caption{Raman spectra in logarithmic scale for three different characteristic particle numbers $\rho = 0.5, 2.5, 10$
           at fixed temperature $k_B T/6J = 0.1$ in three dimensions. $(k_B T/6J)_{f}$ is obtained by fitting the tails of the 
           spectra to Eq.~\ref{eq:3D_slope} \label{fig:3D_sameT_diffrho_log}}
\end{figure}

\begin{figure}[!ht]
 \includegraphics[width=0.85\linewidth,clip=true]{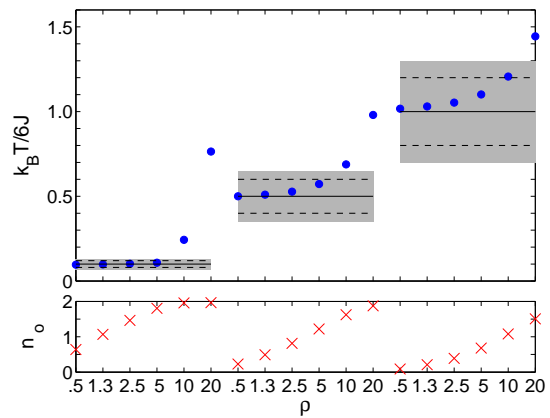}
  \caption{Accuracy of detected temperatures obtained by fitting the tail of each spectrum with Eq.~\ref{eq:3D_slope}.
           Upper panel: detected temperatures are denoted by blue dots while exact temperatures by solid black lines.
           Each shaded region corresponds to a $30\%$ range centered on the exact temperature and contains
           most fitted points. The regions delimited by dashed lines corresponds to a $20\%$ range.
           Lower panel: density at the center of the trap.\label{fig:quality_anal_3D_log}}
\end{figure}

Considering only the tail of the spectrum, a simplified fitting procedure can also be used
to evaluate the temperature of a three dimensional gas. Looking back at the analytical
expression given by Eq.~\ref{eq:rate_free3D},
we see that if $e^{(\tilde{\omega}/2+\mu_o)/k_BT}$ is small \cite{FootNote4}
the Raman signal can be approximated by
\begin{eqnarray}
R^{3D}_{\pi/a}(\tilde{\omega})\propto g_v(\hbar \tilde{\omega}/2) e^{(\hbar\tilde{\omega}/2k_BT)}.
\label{eq:3D_slope}
\end{eqnarray}
As in the two dimensional case, we expect this expression to be accurate for small or even
negative values of $\mu_o$ and small $\hbar\tilde{\omega}/6J\in [-2;2]$. In Fig.~\ref{fig:3D_sameT_diffrho_log},
we apply this simplified fitting method to spectra with various characteristic particle
numbers. For small values of $\rho=0.5, 2.5$, the fit works nicely over a wide range of frequencies.
In contrast for larger values of $\rho$, the range over which the simplified expression
can be fitted becomes very small or nonexistent. In Fig.~\ref{fig:quality_anal_3D_log}, the extracted temperatures
are compared to the input temperatures. As expected the procedure works well for small and intermediate
characteristic particle numbers. In contrast, for large values of $\rho$, this
method overestimates the system temperature.

Finally, to conclude this section, we need to point out that the frequency resolution attainable
in experiments may not be as good as assumed here. Therefore, we checked that the fitting procedure
still works for a reduced frequency resolution by binning the simulated spectra.
For two dimensional systems, the temperature extraction method still works surprisingly well.
The temperature can be determined very accurately even if only a few points are left on
the spectrum ($\hbar\Delta\tilde{\omega}/4J = 0.8$). In three dimensions, the fitting procedure
still works for a frequency resolution of about $\hbar\Delta\tilde{\omega}/6J = 0.4$.

\subsection{Fermions with (moderate) interactions : thermometry from the wings of the density profile}
\label{sec:weakly}

\begin{figure}[!ht]
 \includegraphics[width=0.75\linewidth,clip=true]{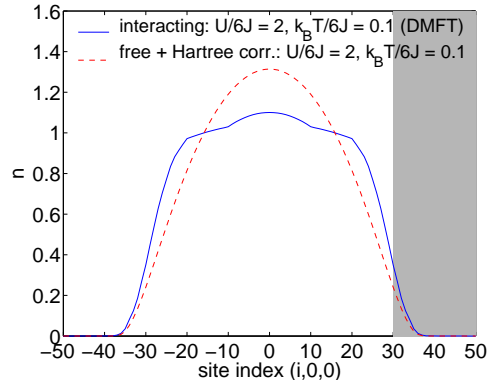}
  \caption{Density cut through ($x$, 0, 0) for an interacting system at $U/6J = 2$, $k_B T/6J = 0.1$ and $\rho = 8.9$.
           Below $n = 0.3$ the interacting and Hartree corrected density profiles agree quite well. The shaded region
           ($x > 30$) corresponds to the probed area, this region contains $0.4\%$ of the total atom number.
           \label{fig:densitycutT0p1}}
\end{figure}

\begin{figure}[!ht]
 \includegraphics[width=0.75\linewidth,clip=true]{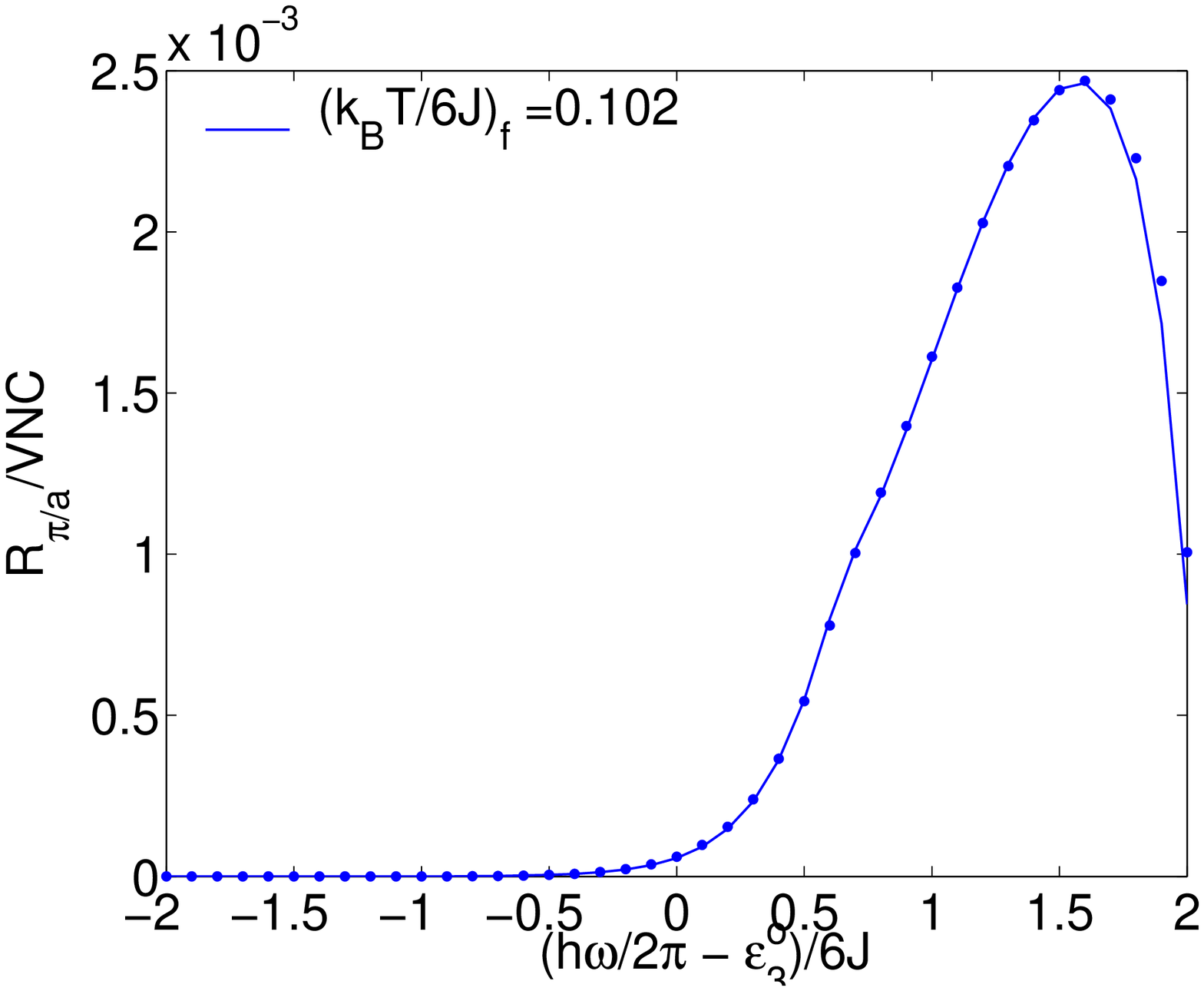}
  \caption{Raman spectrum obtained by only collecting signal from the low density region
           shown on Fig.~\ref{fig:densitycutT0p1}. The system exact temperature is $k_B T/6J = 0.1$ while
           the detected temperature obtained through a fit of Eq.~\ref{eq:ratewi} is $(k_B T/6J)_f = 0.102$.
\label{fig:ramanwiT0p1}}
\end{figure}

\begin{figure}[!ht]
 \includegraphics[width=0.75\linewidth,clip=true]{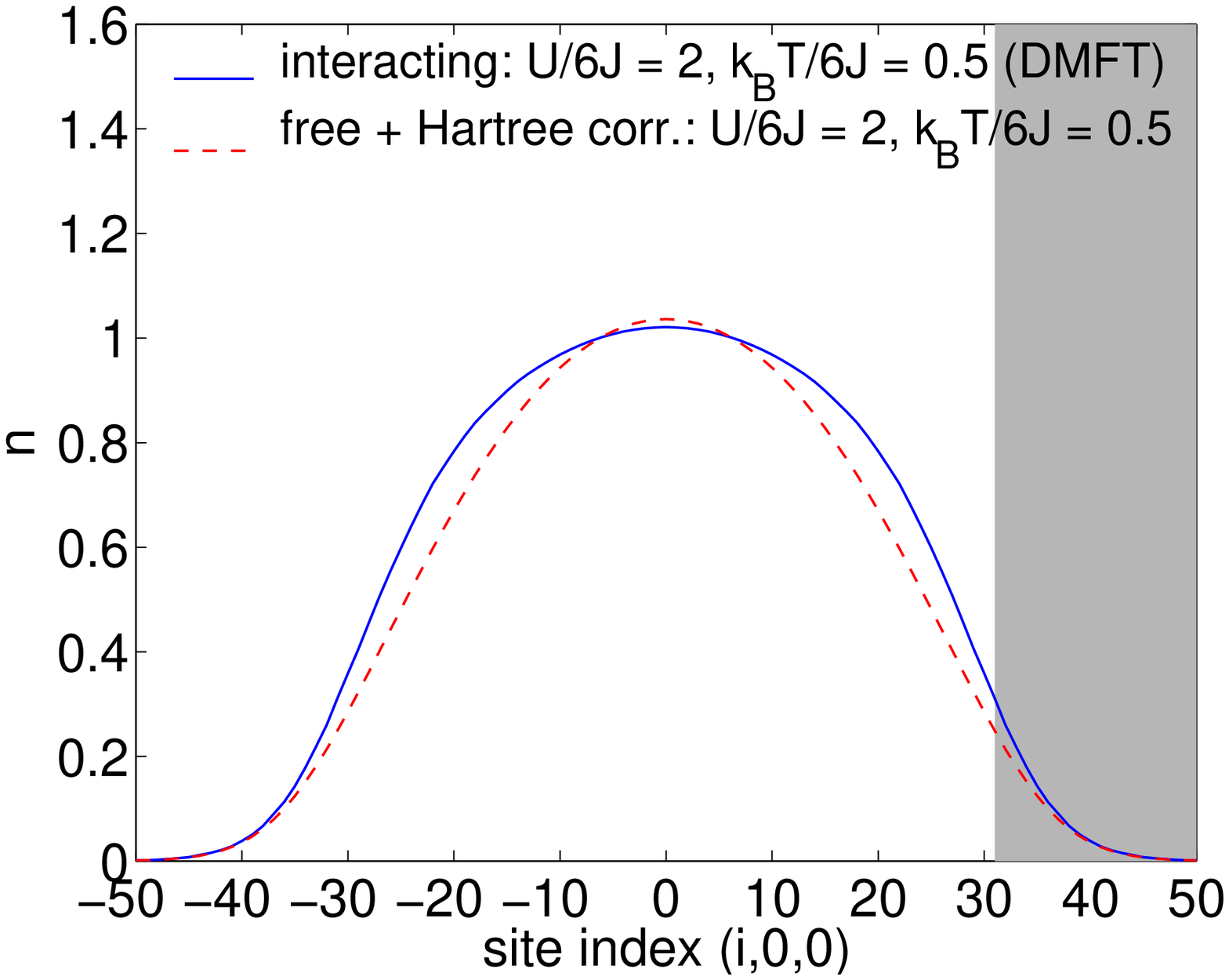}
  \caption{Density cut through ($x$, 0, 0) for an interacting system at $U/6J = 2$, $k_B T/6J = 0.5$ and $\rho = 8.9$.
           Below $n = 0.3$ the interacting and Hartree corrected density profiles agree quite well. The shaded region
           ($x > 31$) corresponds to the probed area, this region contains $1.4\%$ of the total atom number.
           \label{fig:densitycutT0p5}}
\end{figure}

\begin{figure}[!ht]
 \includegraphics[width=0.75\linewidth,clip=true]{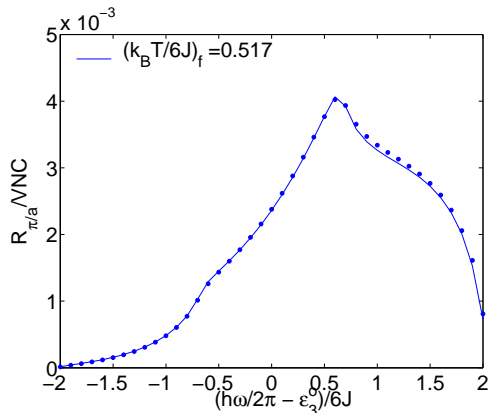}
  \caption{Raman spectrum obtained by only collecting signal from the low density region
           shown on Fig.~\ref{fig:densitycutT0p5}. The system exact temperature is $k_B T/6J = 0.5$ while
           the detected temperature obtained through a fit of Eq.~\ref{eq:ratewi} is $(k_B T/6J)_f = 0.517$.
\label{fig:ramanwiT0p5}}
\end{figure}

When atoms in the $(\ket{1},\ket{1'})$ mixture interact via a finite interaction strength $U$, the structure of the
Raman spectrum changes significantly due to weight redistribution
in the spectral function  (cf.~Sec.~\ref{sec:Mott}). However, up to intermediate interaction strengths, the low density region on the
periphery of the trap is still well described by a system of non-interacting fermions.
Thus, this region can be used to extract the gas temperature assuming the system is in thermal equilibrium. 
The experimental feasibility of the detection of a small boundary region has been shown using radio-frequency 
spectroscopy for an imbalance Fermi mixture and has been used to detect the temperature in the absence of 
an optical lattice potential \cite{ShinKetterle2008}. In contrast to our proposal, the boundary region in 
that case was only occupied by the majority component so the Fermi gas in the wings was clearly non-interacting.
In Fig.~\ref{fig:densitycutT0p1} and \ref{fig:densitycutT0p5}, we show density profiles, obtained from
dynamical mean-field calculations, for two three-dimensional interacting systems.
We compare these profiles to those calculated by using local density approximation and the simple
Hartree approximation for the relation between the density $n$ and
chemical potential $\mu$. The Hartree approximation simply amounts to inverting the relation:
$\mu = \mu_\text{U=0}(n) + \frac{U}{2}\,n$, where $\mu_\text{U=0}(n)$ is the chemical
potential of the free system for a given density $n$.
As one can see from these figures, the interacting and Hartree-approximated profiles agree quite
well for $n < 0.3$ whereas, for larger densities, the profiles are considerably different.
Hence, by only probing the region at the periphery of the trap, we can detect the
system temperature as the atoms at these locations are still described by a
quasi non-interacting model. The validity of this approximation will be further evidenced in Sec.~\ref{sec:Mott}.
Two examples of Raman spectra obtained by
collecting Raman signal coming from one of the six ``semi-spherical'' regions of low density are
shown in Fig.~\ref{fig:ramanwiT0p1} and \ref{fig:ramanwiT0p5}. These Raman spectra are simulated
using Eq.~\ref{eq:lattice} where the sum over positions is limited to one of the six low density regions
and the central chemical potential, $\mu_o$, is the one setting the right atom number
in the interacting system. To show that the temperature can still be well detected
in this limit, we fit these spectra using the continuum Raman expression:
\begin{eqnarray}
\frac{R_{\pi/a}(\tilde{\omega})}{N} &\propto& \frac{g_v(\hbar\tilde{\omega}/2)}{\rho}
\int^{\frac{\pi}{2}}_0\int^{\infty}_{\frac{\alpha}{\cos^2\phi}}
d x \, \sin\phi \, d \phi \nonumber \\
&&~~~~~~ \times \frac{\sqrt{x}}{1+e^{-(\hbar\tilde{\omega}/2+\mu_o-x)/(k_BT)}}.
\label{eq:ratewi}
\end{eqnarray}
In this expression, where the spatial integral is limited to the probed region, there are
only three fitting parameters: the temperature, the central chemical potential and
$\alpha$, a parameter related to the size of the probed region \cite{FootNote5}.
As shown in Fig.~\ref{fig:ramanwiT0p1} and \ref{fig:ramanwiT0p5},  the system temperature can be measured
successfully using this procedure.

\section{Spectra for interacting fermions: from strongly correlated Fermi liquids to Mott insulators}
\label{sec:Mott}

In this section we discuss the structure of the Raman spectrum in different strongly-correlated states.
We present general considerations based on the separate spectral contributions of quasiparticle
excitations and of incoherent high-energy excitations. These considerations are illustrated by
explicit calculations for the Hubbard model with repulsive interactions, treated in the framework
of dynamical mean-field theory (DMFT) \cite{GeorgesRozenberg1996}. The DMFT calculations are performed using the numerical renormalization group
method as an impurity solver \cite{BullaPruschke2008}. The weakly correlated regime, strongly correlated Fermi liquid and
Mott insulating regimes are discussed.

For simplicity, we focus on a homogeneous system, corresponding to Raman spectroscopy being
performed in a local manner \cite{DaoGeorges2009} and probing 
deep inside the bulk of a certain quantum state to avoid the influence of a neighboring state with different character \cite{HelmesRosch2008}. We also restrict our discussion to the paramagnetic phase,
and the calculations are performed at zero temperature (although some qualitative remarks will be made on finite temperature effects).
The possible use of Raman spectroscopy to detect and investigate the magnetically ordered phase is left
for future work.

We focus in this section on zero momentum transfer $\bq=0$. This is in contrast to the
previous section, in which we used $\bq=\pi/a$ in order to spread the signal as much as possible
to probe thermally excited states. Here, on the contrary, we want to separate and 
resolve the different spectral features (e.g. quasiparticles and
Hubbard bands) as well as possible, and for this $\bq=0$ is more favorable \cite{FootNote6}.
We focus both on the momentum-resolved spectrum (i.e. after time of flight) and on the momentum integrated
signal (the latter being easier to achieve experimentally in the lattice) related to the
spectral function $A(\bk,\freq)$ \cite{ImadaTokura1998} by:
\begin{eqnarray}
R_{\bq=0}(\bk,\omega) &=& C~n_F(\varepsilon_3^o-\hbar\omega+\ek-\mu) \nonumber \\
&& ~~~~~~\times A({\bf k},\varepsilon_3^o-\hbar\omega+\ek-\mu),
\label{eq:rate-kresolved}
\end{eqnarray}
\begin{eqnarray}
R_{\bq=0}(\omega) &=& C~\int d\bk\,n_F(\varepsilon_3^o-\hbar\omega+\ek-\mu) \nonumber \\
&& ~~~~~~~~~ \times A({\bf k},\varepsilon_3^o-\hbar\omega+\ek-\mu).
\label{eq:rate-kint}
\end{eqnarray}
We recall that, in these expressions, $\varepsilon_3^o+\ek$ is the dispersion of
the outcoupled state $|3\rangle$, while $\mu$ is the chemical potential of the interacting
$(|1\rangle,|1'\rangle)$ mixture.
From Eq.~\ref{eq:rate} the prefactor reads $C=2\pi|\Omega_e|^2/\hbar$ (we note that
the Wannier matrix element $W_{\bk}^{\bq=0}=1$ assuming the same lattice potential for state $\ket{1}$ 
and $\ket{3}$).

When specializing to DMFT calculations, the self-energy only depends on
frequency, so that the spectral function
$\pi A(\bk,\freq)\equiv -\rm{Im}\{1/[\freq+\mu-\ek-\Sigma(\freq+i0^+)]\}$ depends on
momentum through $\ek$ only. In this case, the momentum integration can be replaced by
an integration over the density of states, $g_v(\varepsilon)$, associated with the
dispersion $\ek$ (for simplicity, the DMFT calculations presented below will be performed
for a semi-circular density of states):
\begin{eqnarray}
R_{\bq=0}(\omega) &=&  C~V~\int d\varepsilon~g_v(\varepsilon)\,n_F(\varepsilon_3^o-\hbar\omega+\varepsilon-\mu) \nonumber \\
&& ~~~ \times A({\bf k},\varepsilon_3^o-\hbar\omega+\varepsilon-\mu).
\label{eq:rate-kint-DMFT}
\end{eqnarray}
At $T=0$, the Fermi function in these expressions limits the integration domain to
momenta such that $\varepsilon < \mu + \hbar\omega - \varepsilon_3^o$.

We note that these spectra obey the following sum-rules, valid at arbitrary
temperature $T$:
\begin{equation}
\int d\omega R_{\bq=0}(\bk,\omega) = C\,n_1(\bk)\,\,\,,\,\,\,
\int d\omega R_{\bq=0}(\omega) = C\,N/2.
\end{equation}
Hence, the total intensity of the momentum-resolved signal is proportional
to the momentum distribution $n_1(\bk)\equiv \langle c^\dagger_{\bk,1} c_{\bk,1}\rangle$
of particles of type $\ket{1}$ in the system, while the total intensity of the
momentum-integrated signal is proportional to the total number of particles $N/2$ in state $\ket{1}$. In these expressions, the
frequency integration is over the whole range of frequencies where the signal is non-zero (this range is
bounded from below, as shown later).

\begin{figure}[!ht]
  \includegraphics[width=1\linewidth,clip=true]{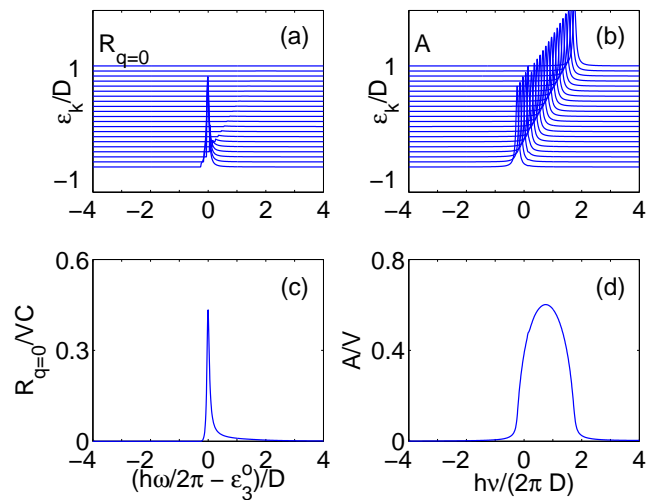}
    \caption{Spectra for a low density liquid. $\mu/D = -0.75$, $U/D = 3.5$, $n \approx 0.18 $. (a) Momentum-resolved Raman spectrum (in arbitrary units). 
            (b) Momentum-resolved spectral function (in arbitrary units). (c) Momentum-integrated Raman spectrum. (d) Momentum-integrated spectral function.
\label{fig:mum0p75U3p5}}
\end{figure}

\begin{figure}[!ht]
 \includegraphics[width=1\linewidth,clip=true]{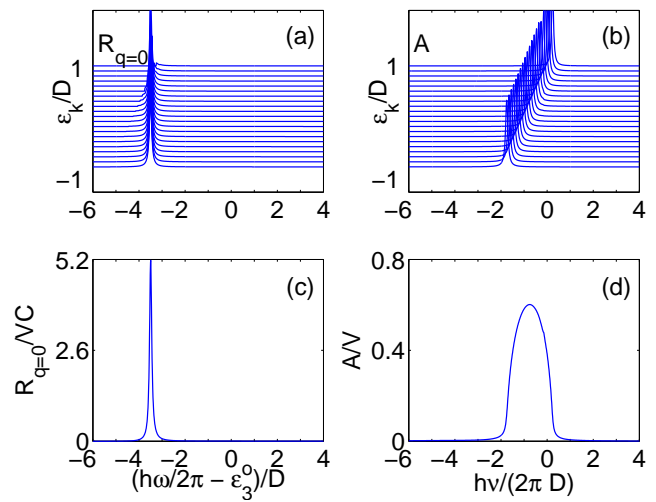}
    \caption{Spectra for a high density liquid. $\mu/D = 4.25$, $U/D = 3.5$, $n \approx 1.82$. 
             (a) Momentum-resolved Raman spectrum (in arbitrary units). (b) Momentum-resolved spectral
             function (in arbitrary units). (c) Momentum-integrated Raman spectrum. (d) Momentum-integrated spectral function.
\label{fig:mu4p25U3p5}}
\end{figure}

%
Let us first discuss the shape of the Raman spectrum in the simple case of a
non-interacting system for which $A(\bk,\freq)=\delta(\freq+\mu-\ek)$. We obtain in this case:
\begin{eqnarray}
&&R^{U=0}_{\bq=0}(\bk,\omega) = C~\delta(\freqR)\,n_F(\ek-\mu),\nonumber \\
&&R^{U=0}_{\bq=0}(\omega) = C~\frac{N}{2} \delta(\freqR).
\end{eqnarray}
Hence, at $\bq=0$ and in the absence of interactions, Raman transitions only exists at
the frequency $\hbar\omega=\varepsilon_3^o$. This is due to the assumption that the
dispersions for the atoms in the $(\ket{1},\ket{1'})$ mixture and in the outcoupled state $\ket{3}$ are the same 
(cf. Fig.~\ref{fig:hom_bands}). 
At $T=0$, the momentum-resolved signal is
non-zero only for momenta inside the Fermi surface $\ek<\mu$ because this spectroscopy
probes only occupied states. At $T\neq 0$, the signal extends beyond the Fermi surface because of
thermal broadening according to the Fermi function.
%
These simple considerations are nicely illustrated
by the spectra displayed in Figs.~\ref{fig:mum0p75U3p5} and \ref{fig:mu4p25U3p5}.
These results correspond to the
Hubbard model with a very low density of particles per site ($n\approx 0.18$, dilute system) and
a very high density of particles per site ($n\approx 1.82$, or low density of holes in a
band insulator), respectively. Despite the fact that these DMFT calculations were done
for a rather high value of $U/D=3.5$, the system is in effect weakly correlated because
the density of particles (or holes) is small. Here $D$ is the half bandwidth. This is clearly seen from 
the displayed momentum-resolved spectral functions (Figs.~\ref{fig:mum0p75U3p5}, \ref{fig:mu4p25U3p5} (b))
as they are weakly modified as compared to the non-interacting case. Both spectra show a very sharp peak
which disperses essentially according to the free dispersion $\varepsilon_{\bf k}$ (only a shift in position is seen).
Hence, the Raman spectra are closely following
the non-interacting behavior: the momentum-integrated spectra (Figs.~\ref{fig:mum0p75U3p5}, \ref{fig:mu4p25U3p5} (c))
are sharply peaked, 
the momentum-resolved Raman spectra (Figs.~\ref{fig:mum0p75U3p5}, \ref{fig:mu4p25U3p5} (a)) have very little
momentum dispersion (in contrast to the spectral function itself), and are suppressed for momenta outside
the Fermi surface. Due to this narrow momentum dispersion a very sharp peak occurs in the momentum-integrated Raman-spectra (Figs.~\ref{fig:mum0p75U3p5}, \ref{fig:mu4p25U3p5} (c)) (the peak position will be discussed later). Let us emphasize that these findings further support the 
detection scheme for interacting particles presented in Sec. \ref{sec:weakly} which relies on the assumption that in the low density 
regions the spectral function behaves like the one of non-interacting particles.
%

We now turn to spectra in which effects of strong correlations become more pronounced. In order to
discuss these spectra on a general basis, we can separate the spectral function into
a contribution from quasiparticles and a contribution from high-energy incoherent excitations:
\begin{equation}
A(\bk,\freq) = A_{\rm{qp}}(\bk,\freq)\,+\,A_{\rm{inc}}(\bk,\freq).
\label{eq:dec-A}
\end{equation}
The quasiparticle contribution can be appropriately described, at low excitation energies and
close to the Fermi surface, by a sharply peaked Lorentzian:
\begin{equation}
A_{\rm{qp}}(\bk,\freq) \simeq \frac{Z_\bk}{\pi}\,
\frac{\Gamma_\bk}{[\freq-(\ekqp-\mu)]^2+\Gamma_\bk^2}.
\label{eq:Aqp}
\end{equation}
In this expression, $\ekqp$ is the dispersion relation of quasiparticles, $\gammak$ is their
inverse lifetime and $\Zk$ the spectral weight associated with the contribution of quasiparticles
to the total spectrum of single-particle excitations \cite{FootNote7}.
In a Fermi liquid,
the quasiparticle excitations become long-lived coherent excitations as
the Fermi surface is approached, corresponding to a sharp peak with
width $\gammak \propto (\ekqp-\mu)^2 \sim (\bk-\bk_F)^2$.

\begin{figure}[!ht]
 \includegraphics[width=1\linewidth,clip=true]{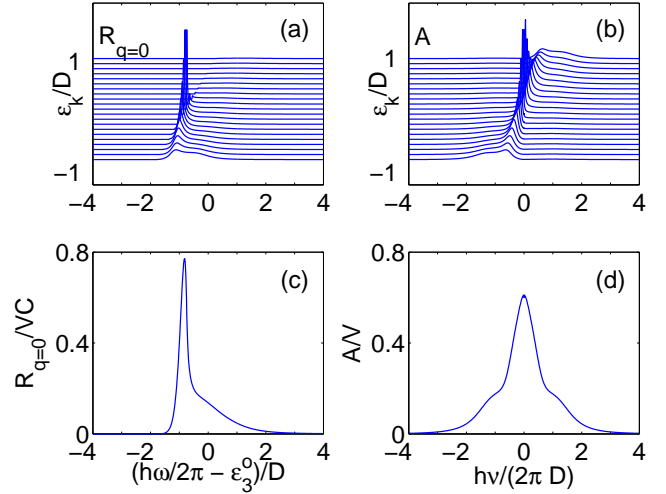}
    \caption{Spectra for a Fermi liquid with moderate correlations. $\mu/D = 0.75$, $U/D = 1.5$, $ n = 1$. 
             (a) Momentum-resolved Raman spectrum (in arbitrary units). (b) Momentum-resolved spectral
             function (in arbitrary units). (c) Momentum-integrated Raman spectrum. (d) Momentum-integrated spectral function.
\label{fig:mu0p75U1p5}}
\end{figure}

To illustrate how quasiparticles contribute to the Raman spectrum, we display
in Fig.~\ref{fig:mu0p75U1p5}
the results of a DMFT calculation for the half-filled Hubbard model at $U/D=1.5$, which
corresponds to a Fermi liquid in the intermediate correlation regime.
The momentum-resolved spectral function (Fig.~\ref{fig:mu0p75U1p5}~(b)) clearly displays a quasiparticle
peak. This quasiparticle peak becomes sharp as the Fermi surface is reached (corresponding here to
$\varepsilon_{\bk_\text{F}}=0$), while for momenta far from the Fermi surface only a
broader incoherent contribution is seen. The momentum-resolved Raman spectrum (Fig.~\ref{fig:mu0p75U1p5}~(a)) 
shows the same features below the Fermi level. However, the dispersion of the quasi-particle peak close to ${\bf k}_\text{F}$ 
and the incoherent contribution far from the Fermi surface behave differently than in the spectral function.
The momentum-integrated
Raman spectrum (Fig.~\ref{fig:mu0p75U1p5}~(c)) has a well-marked peak corresponding to quasiparticle contribution to
the density of states, and a broder hump corresponding to incoherent excitations.
In order to understand better these spectral features, we note that the contribution of
quasiparticles to the momentum-resolved Raman spectrum reads, using (\ref{eq:Aqp}) into
(\ref{eq:rate-kresolved}):
\begin{eqnarray}
R^{\rm{qp}}_{\bq=0}(\bk,\omega) &\simeq& C~n_F(\ekqp-\mu)\,\frac{Z_\bk}{\pi}\ \nonumber \\
&& \times~\frac{\Gamma_\bk}{[\freqR-(\ek-\ekqp)]^2+\Gamma_\bk^2}.
\label{eq:qp_spectrum}
\end{eqnarray}
From this expression, it is clear that the quasiparticle peak in the momentum-resolved
Raman spectrum disperses according to:
$(\hbar\omega -\varepsilon_3^o)^{\rm{qp}}_{\bf k}=\ek-\ekqp\sim ({\bf v}_F-{\bf v}_F^{\rm{qp}})\cdot(\bk-\bk_F)+\dots$.
The last expression is valid for momenta near the Fermi surface and involves
the {\it difference} between the actual Fermi velocity, ${\bf v}_F^{\rm{qp}}$, in the presence of interactions
(related to the effective mass) and the bare Fermi velocity, ${\bf v}_F$. 
Indeed, the peak in the Raman signal is less dispersive (Fig.~\ref{fig:mu0p75U1p5}) than the
one in the spectral function (dispersing as ${\bf v}_F^{\rm{qp}}\cdot({\bf k}-{\bf k}_F)$). In practice, since the dispersion
$\ek$ of the outcoupled band is known, it shall be possible to extract directly $\ekqp$ from
the Raman signal by plotting it as a function of $\freqR-\ek$
(as done in \cite{StewartJin2008} in the continuum).
As the quasiparticle peak becomes very sharp near the Fermi surface $\bk\simeq \bk_F$,
those momenta dominate the momentum-integration (at least for lattices with a non-singular $g_v(\varepsilon)$).
Hence, the momentum-integrated spectrum (Fig.~\ref{fig:mu0p75U1p5}~(c)) has a quasiparticle peak located at
$(\hbar\omega-\varepsilon_3^o)^{\rm{qp}} = \langle\varepsilon_{\bk_F}\rangle -
\mu \simeq \mu_{U=0}-\mu$. In the first expression, $\langle\varepsilon_{\bk_F}\rangle$
corresponds to a Fermi surface average. The second expression is valid when the
Fermi surface is only mildly deformed by interactions, so that the Luttinger theorem
(conservation of Fermi surface volume) implies that $\varepsilon_{\bk_F}=\mu_{U=0}$, where $\mu_{U=0}$ is the
chemical potential of the non-interacting system at the same density. This analysis
accounts well for the location of the peak (at $\sim -\mu$) in the spectrum of
Fig.~\ref{fig:mu0p75U1p5}~(c)  (which corresponds to half-filling, so that $\mu_{U=0}=0$,
while $\mu=U/2=0.75\,D$). The onset of Raman absorption in the momentum integrated
spectrum at $T=0$ corresponds to the restriction due to the Fermi function
$\freqR>\ek-\mu$ and hence corresponds to a threshold frequency:
$(\hbar\omega-\varepsilon_3^o)^{\rm{th}}=-D-\mu$, again well obeyed in Fig.~\ref{fig:mu0p75U1p5}~(c).
We note that this absorption threshold corresponds to the transfer of states from the bottom of the band with $\bk=0$
(i.e.~occupied states well below the Fermi surface), as it is known from
radio-frequency spectroscopy \cite{KetterleZwierlein2007,BlochZwerger2008}.
%

\begin{figure}[!ht]
 \includegraphics[width=1\linewidth,clip=true]{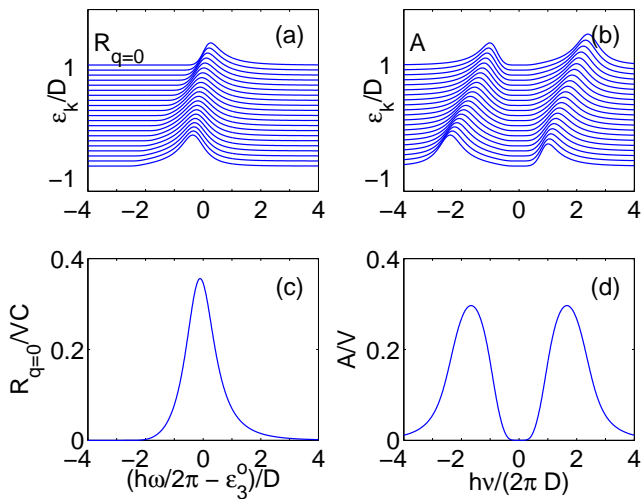}
    \caption{Spectra for a Mott insulator. $\mu/D = 1.75$, $U/D = 3.5$, $n = 1$. (a) Momentum-resolved Raman spectrum (in arbitrary units). 
            (b) Momentum-resolved spectral function (in arbitrary units). (c) Momentum-integrated Raman spectrum. 
            (d) Momentum-integrated spectral function.
\label{fig:mu1p75U3p5}}
\end{figure}

Having discussed a Fermi liquid in the regime of intermediate correlations, we
turn to the opposite limit of a very strongly correlated system: a Mott insulator, as
realized e.g. in the Hubbard model at half-filling and for large interaction strength
($U/D=3.5$ in Fig.~\ref{fig:mu1p75U3p5}).
There, quasiparticles are absent and the high-energy incoherent
excitations correspond to the Hubbard ``bands''. The lower (upper) Hubbard band (LHB, resp. UHB)
corresponds to the process of removing (adding) an atom on a singly occupied site.
This corresponds to two peaks in the spectral function (Fig.~\ref{fig:mu1p75U3p5}~(b) and (d)) at
$\freq_{\bk}^{\rm{LHB}}<0$ and $\freq_{\bk}^{\rm{UHB}}>0$, separated by the Mott gap $\Delta_g$.
Since the excitation energy from the ground-state for removing a particle is $\mu$, the
lower Hubbard band $\freq_{\bk}^{\rm{LHB}}$ is centered at $\sim -\mu$.
This band disperses over a bandwidth of order $D$ with a width $\Gamma_\bk^{\rm{LHB}}$ of order $D$ itself. 
Hence, the excitation is  `incoherent' in nature (except at momenta near the top of the band where the
width is smaller, of order $D^2/U$). Similar considerations apply to the upper Hubbard band (centered at
$\sim U-\mu$). The total (momentum-integrated) weight of the lower Hubbard band in the spectral
function is proportional to $n/2$, while that
of the upper Hubbard band is proportional to $1-n/2$.

At $T=0$, the lower Hubbard band is fully visible in the Raman spectrum, as seen on Fig.~\ref{fig:mu1p75U3p5}~(a) and (c) 
(there, $-\mu=-U/2=-1.75\,D$). This lower band in the momentum-resolved Raman spectrum is located at:
$(\hbar\omega-\varepsilon_3^o)_\bk^{\rm{LHB}}=\ek-\mu-\freq_{\bk}^{\rm{LHB}}$.
Since $\freq_{\bk}^{\rm{LHB}}<0$, the lower Hubbard band is apparent for all momenta (in contrast to
a quasiparticle peak which is suppressed as the Fermi surface is crossed).
As the momentum integration is dominated by $\ek=0$ and $\freq_{\bk}^{\rm{LHB}}$ is centered at
$-\mu$, the lower Hubbard band results in a peak in the momentum-integrated spectrum
located at $(\hbar\omega-\varepsilon_3^o)^{\rm{LHB}}\simeq\,-\mu-(-\mu)=0$, as clear on
Fig.~\ref{fig:mu1p75U3p5}~(c).
The threshold for Raman absorption corresponds to:
$(\freqR)^{\rm{th}}=\rm{Min}_{\bk} [\ek-\freq_{\bk}^{\rm{LHB}}]-\mu$. The minimum is usually
realized for $\ek=-D$, so that:
$(\freqR)^{\rm{th}}=-D-\freq_{\rm{top}}^{\rm{LHB}}-\mu$. At half-filling this
reads: $(\freqR)^{\rm{th}}=-D+\Delta_g/2-U/2$, with $\Delta_g$ the Mott gap.
For temperatures comparable or higher than the Mott gap, the upper Hubbard band will
become visible in Raman spectra, at a location: $(\freqR)^{\rm{UHB}}\simeq -U$. Concretely,
imaging the integrated or full Hubbard bands would be very useful as it would not only give information on these incoherent excitations themselves, but as well provide a novel method to
extract the interacting strength, $U$, and the gap size, $\Delta_g$.

%
%

\begin{figure}[!ht]
 \includegraphics[width=1\linewidth,clip=true]{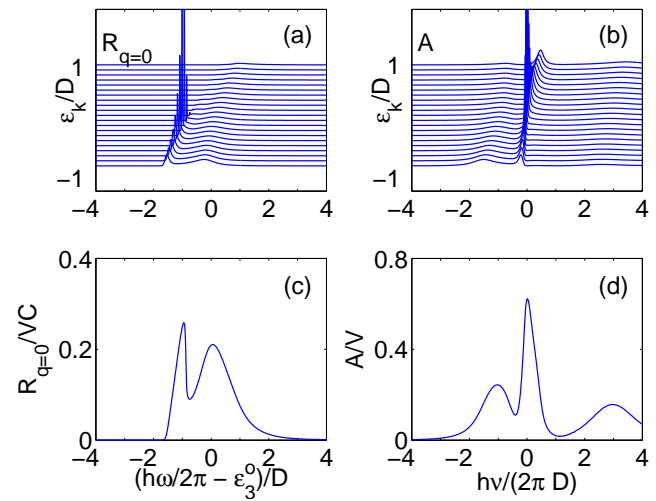}
    \caption{Spectra for a strongly correlated Fermi liquid. $\mu/D = 0.75$, $U/D = 3.5$, $n \approx 0.85$. 
             (a) Momentum-resolved Raman spectrum (in arbitrary units). (b) Momentum-resolved spectral
             function (in arbitrary units). (c) Momentum-integrated Raman spectrum. (d) Momentum-integrated spectral function.
\label{fig:mu0p75U3p5}}
\end{figure}

Finally, we display results for a strongly correlated Fermi liquid with a
spectral function that displays simultaneously a central peak of
quasiparticle excitations, as well as lower and upper Hubbard bands
(Fig.~\ref{fig:mu0p75U3p5}, corresponding to
a rather large coupling $U/D=3.5$ with $n=0.85$, i.e. to a strongly correlated Fermi
liquid).
The Raman spectra reveal both types of excitations, which also lead to two distinct
features in the momentum-integrated Raman spectrum (Fig.~\ref{fig:mu0p75U3p5}~(c))
at frequencies expected from the analysis above.
We note that these two features will in general have very different temperature dependences.
As the temperature is raised, the quasiparticle peak will be suppressed when temperature
exceeds the quasiparticle coherence temperature, of order $Z_{\bk_F}D$. In contrast, the
lower Hubbard band will start losing weight (and the upper Hubbard band will start appearing) only at a
higher temperature scale comparable to the gap scale.

In summary, Raman spectroscopy is a useful probe to explore various possible regimes of
correlations. Broad Hubbard bands are seen in the incompressible Mott regime, while the additional observation 
of a quasiparticle peak at low temperature signals the formation of a strongly correlated Fermi liquid.

\section{Conclusion}
In this work, we demonstrated that Raman spectroscopy is
a versatile probe that can be used to measure the temperature of Fermi
gases confined to optical lattices and to identify various
signatures of strongly correlated fermionic phases. The proposed
detection scheme, implementable
with present technology, relies on transferring a portion of the atoms
stored into the optical lattice potential to a third hyperfine state. This
Raman rate can in principle be both resolved in frequency and
momentum. We showed that momentum resolution is not required
to accurately measure the temperature of free and weakly interacting
fermionic atoms loaded into an optical lattice and that the detection
can either be done locally or globally. We also demonstrated that
detecting several features of strongly correlated liquids and Mott
insulators such as quasiparticle peaks and Hubbard bands can be done
using the same scheme without knowledge of the atom momentum.
However, in the future, if momentum resolution is experimentally
achieved in a lattice, momentum resolved Raman rate could provide
valuable information on the level of correlation of fermionic cold
atom systems. Finally, we would like to point out that Raman
spectroscopy can even be used to cool down fermionic atoms confined to
an optical lattice as explained in \cite{GriessnerZoller2006}.

\acknowledgments We are grateful to I. Bloch, I. Carusotto, J. Dalibard, M. K\"ohl, L. Perfetti, C. Salomon and the Quantum Optics group
of ETH Z{\"u}rich for stimulating discussions. We acknowledge support from the Triangle de la Physique,
the Agence Nationale de la Recherche (under contracts FABIOLA and FAMOUS),
the DARPA-OLE program, the Fonds Qu\'eb\'ecois de la Recherche sur la Nature et les Technologies, and ANPCyT Grant No 482/06.


\begin{thebibliography}{29}
\bibitem{FootNote1}
\bibinfo{author}{Radio-frequency spectroscopy is also possible for the cases of vanishing momentum transfer.}
\bibitem{FootNote2}
\bibinfo{author}{In the case of a single frequency, the linear response is valid for a more restrictive time interval.}
\bibitem{FootNote4}
\bibinfo{author}{If this condition is fulfilled for $\mu_o$ it also is for the whole integration range.}
\bibitem{FootNote5}
\bibinfo{author}{$\alpha = V_T r^2_\text{min}$ where $r_\text{min}$ is the radial position at which $n=0.3$.}.
\bibitem{FootNote6}
\bibinfo{author}{$\bq=0$ can also be realized using radio-frequency spectroscopy.}
\bibitem{FootNote7}
\bibinfo{author}{These quantities are related to the self-energy by the standard expressions:
$\Zk=1/(1-\partial\Sigma/\partial(\hbar\nu)|_{\nu=0,\bk})\,,\ekqp=\Zk\,[\ek+(\bk-\bk_F)\cdot\nabla_\bk\Sigma]\,,
\gammak=\Zk|\rm{Im}\{\Sigma(\bk,\ekqp-\mu)\}|$.}
\end{thebibliography}

\end{document}